\documentclass[bibyear]{aa} % if the references are not structured 
\usepackage{graphicx}
\usepackage{graphicx}
\usepackage{txfonts}
\usepackage{url}
\usepackage{hyperref}

\usepackage{natbib}
\usepackage{color}

\newcommand{\uvot}{\textit{Swift}/UVOT}
\newcommand{\xrt}{\textit{Swift}/XRT}

\newcommand{\nus}{NuSTAR}

\newcommand{\po}{power-law}

\newcommand{\lp}{logparabola}

\newcommand{\nh}{N$_H$}
\newcommand{\obj}{BL\,Lacertae}

\begin{document} 

%\linenumbers            % uncomment for submission
%\linenumbersep 3pt\relax                % uncomment for submission
%\def\linenumberfont{\normalfont\tiny\sffamily}          % uncomment for submission

   \title{Exceptional X-ray activity in BL Lacertae}

\author{Alicja Wierzcholska
          \inst{1,2}
          \and
          Stefan Wagner\inst{2}
          }

   \institute{Insitute of Nuclear Physics, Polish Academy of Sciences, ul. Radzikowskiego 152, 31-342 Krak\'{o}w, Poland \\
              \email{alicja.wierzcholska@ifj.edu.pl}
         \and Landessternwarte, Universit\"at Heidelberg, K\"onigstuhl 12, D 69117 Heidelberg, Germany
              \\
         }

\abstract{
BL\,Lacertae is a unique blazar for which the X-ray band can cover either the synchrotron or the inverse Compton, or both parts of the broadband spectral energy distribution. 
In the latter case, when the spectral upturn is located in the X-ray range, it allows contemporaneous study of the low- and high-energy ends of the electron distribution function.

In this work, we study spectral and temporal variability using X-ray and optical observations of the blazar performed with the \textit{Neil Gehrels Swift} Observatory from 2020 to 2023. 
The large set of observational data reveals intensive flaring activity, accompanied by spectral changes in both spectral branches.

We conclude that the low-energy and high-energy ends of the particle distribution function are characterised by similar variability scales. 
Additionally, the hard X-ray observations of \obj\ performed with the Nuclear Spectroscopic Telescope Array (\textit{\nus)}\  confirm a concave spectral curvature for some epochs of the blazar activity and reveal that it can be shifted up to energies of as high as 8\,keV.

The time-resolved spectral analysis allows us to disentangle X-ray spectral variability features of the synchrotron from inverse Compton components.
Despite significant variability of both spectral components, we find only small changes in the  position of the spectral upturn.

The different slopes and shapes of the X-ray spectrum of \obj\ demonstrate that the classification of this source is not constant, and \obj\ can exhibit features of either high-, intermediate-, or low-energy peaked blazar in different epochs of observation. 
This also indicates that the spectral upturn for this blazar can be located not only  in the X-ray range of 0.3-10\,keV, but also at lower or higher energies.

   } 
 
   \keywords{galaxies: active -- galaxies: jets -- galaxies: individual: BL Lacertae -- $\gamma$ rays: galaxies
               }

   \authorrunning{}
   \titlerunning{}         
               
   \maketitle

\section{Introduction}
Blazars, a class of active galactic nuclei (AGN) composed of BL Lacertae-(BL Lac)-type objects and flat-spectrum radio quasars (FSRQs), include sources characterised by a polarised and highly variable non-thermal emission. 
The emission of blazars is believed to be dominated by non-thermal radiation originating from a jet pointing towards the observer \citep{begelman84}.
The emission of blazars is observed at all frequencies, from radio to high- and very high-energy $\gamma$ rays \citep[e.g.][]{Wagner2009, Vercellone,  Aharonian13_0301, Abramowski2014, Wierzcholska_atom, Wierzcholska_host}.
The broadband spectral energy distribution (SED) of blazars has a typical structure with two prominent bumps. 
The low-energy bump is attributed to synchrotron radiation of relativistic electrons from the jet,  while the origin of the high-energy bump is still under debate \citep[see e.g.][]{Maraschi92, Kirk98, Aharonian00, Atoyan03, Mucke13, Bottcher13}.
The most common explanation for this double structure is the synchrotron self-Compton model \citep[SSC; e.g.][]{Maraschi92, Kirk98}, where the origin of the high-energy bump is attributed to inverse Compton radiation scattered by the same population of relativistic electrons.
Alternatively, the high-energy peak can be interpreted within the context of hadronic models. In these scenarios, the high-energy radiation originates from processes such as
a proton synchrotron emission, synchrotron, and Compton emission from secondary decay products of charged pions, or $\pi_{0}$ decay \citep[see e.g.][]{Aharonian00, Mucke13, Bottcher13}.

Among BL Lac-type blazars, high-, intermediate-, and low-energy peaked objects can be distinguished, and are known as HBL, IBL, and LBL objects, respectively \citep[see e.g.][]{padovani95,fossati98,Abdo2010}. 
Correspondingly, the X-ray spectrum of the different sub-classes of BL Lac objects covers different components of the broadband SED.
For HBL-type blazars, in the broadband SED, the X-ray spectrum is a part of the synchrotron domain.
For LBL-type blazars, the X-ray spectrum belongs to the inverse Compton bump.
In the case of IBL-type sources, the X-ray emission usually covers either the synchrotron component or the inverse Compton component, or both. 
The spectral upturn in the X-ray domain has
been reported for several IBL-type sources, such as  BL~Lacertae \citep{Tanihata00,Ravasio02,Donato05, Wierzcholska_swift}, W~Comae \citep{Tagliaferri00, Donato05, Wierzcholska_swift}, S5~0716+71 \citep{Cappi94,Giommi99,Tagliaferri03,Donato05,Ferrero06, Wierzcholska_s5_2015,  Wierzcholska_swift, Wierzcholska_S5nustar}, AO~0235+16 \citep{Raiteri06}, OQ~5310 \citep{Tagliaferri03}, 3C~66A \citep{Donato05, Wierzcholska_swift}, and 4C\,+21.35\citep{Wierzcholska_swift}.

Studies of X-ray observations of different blazars have revealed strong variability in this energy regime, with different amplitudes of variability and on different timescales
 \citep[e.g.][]{Sembay93, Tanihata00, Sembay02, Zhang02, Zhang2005}.
Also, for several HBL blazars, a so-called `harder-when-brighter' behaviour manifests as hardening of the spectral index with increasing flux level \citep[e.g.][]{Pian98, Zhang2005, Zhang06b, Abramowski2012_2155, Wierzcholska2155_gal}.
This feature is not common to all subclasses of blazars.
In the case of IBL-type blazars in particular, in which X-ray spectra are due to different processes, the variability patterns are expected to be more complex.

BL Lacertae, a blazar at a  redshift of z = 0.069 \citep{Miller_red}, is the prototype of the BL Lacertae class of blazars. 
This particular source is classified as either LBL \citep{Nilsson_bllac} or IBL \citep{2FGL}.
The source has been observed in all energy regimes, starting from radio up to very high-energy $\gamma$ rays, with several multiwavelength campaigns targeting the blazar \citep[e.g.][]{Bertaud69, Madejski99, Villata04, Papadakis07, Villata09, Raiteri13, Agarwal15,  MAGIC19}.
Several X-ray campaigns performed with instruments such as HEAO–1, Einstein, Ginga, ROSAT, ASCA, \textit{Beppo}SAX, RXTE, and \textit{Swift} revealed significant spectral and temporal variability of \obj. 

In the present work, we aim to characterise the spectral and temporal X-ray variability of \obj\ based on recent (2020-2023) \textit{Swift} observations of the blazar. Additional information about the X-ray properties of the blazar is added thanks to \textit{\nus}\ data. 
The paper is organised as follows: Section\,\ref{obs}  describes the data used and our analysis techniques. Sections\,\ref{longlc}-\ref{nus} focus on the description of the spectral and temporal variability of \obj, and we outline our conclusions in Sect.\,\ref{summary}.

\section{Observations and data analysis}\label{obs}

\subsection*{Swift-XRT and Swift-UVOT observations}

The \textit{Neil Gehrels Swift} Observatory \citep[hereafter \textit{Swift};][]{Gehrels04} launched in November 2004 is a multiwavelength space observatory equipped with three instruments:  the Burst Alert Telescope \citep[BAT;][]{Barthelmy05}, the X-ray Telescope \citep[XRT;][]{Burrows05}, and the Ultraviolet/Optical Telescope \citep[UVOT;][]{Roming05}, allowing observations in X-ray, ultraviolet, and optical energy ranges.

Our detailed spectral and temporal analysis focuses on observations collected in the period of 2020-2023,  corresponding to the ObsIDs of 00030720231-00035028001. 
The X-ray data in the energy range of 0.3-10\,keV collected with \xrt\ were analysed using 6.31.1 of the HEASOFT package\footnote{\url{http://heasarc.gsfc.nasa.gov/docs/software/lheasoft}}.
The data were recalibrated using the standard procedure \verb|xrtpipeline|.
For the spectral fitting, \verb|XSPEC| v.12.13.1  was used \citep{Arnaud96}.
All data were binned to have at least one count per bin.

We used version 6.31.1 of the HEASOFT package to analyse the optical and ultraviolet observations of \uvot\ in six bands, namely UVW2 (188\,nm), UVM2 (217\,nm), UVW1 (251\,nm), U (345\,nm), B (439\,nm), and V (544\,nm).
The instrumental magnitudes were calculated using \verb|uvotsource|, including all photons from a circular region with a radius of 5''.
The background was determined from a circular region with a radius of 10'' near the source region not contaminated with the signal from nearby sources. 
The flux conversion factors as provided by \cite{Poole08} were used. 
All data were corrected for the dust absorption using the reddening $E(B-V)$ = 0.2821   as provided by \cite{Schlafly}.
The ratios of the extinction to reddening, $A_{\lambda} / E(B-V),$ for each filter were provided by \cite{Giommi06}.

\subsection*{NuSTAR observations}
The Nuclear Spectroscopic Telescope Array (\textit{NuSTAR}) is a satellite instrument dedicated to observations in the hard X-ray regime up to 79 keV \citep{Harrison13}.
All observations were performed in the \verb|SCIENCE| mode.
The raw data were processed with the \textit{\nus}\ Data Analysis Software package (\verb|NuSTARDAS|, 
released as part of \verb|HEASOFT|~6.31.1) using the standard  \verb|nupipeline| task.  
Instrumental response matrices and effective area files were produced with the
\verb|nuproducts| procedure. 
The spectral analysis was performed for the channels corresponding to the energy band of 3-79\,keV.

\textit{\nus}\ observed \obj\ seven times.
All these observations are listed in Table\ref{table:nustardata} and marked as N1-N7.
We note here that \obj\ was observed twice with \textit{\nus}\  in 2012; however, due to its very short duration, the first observation is neglected in these studies. 
Table \ref{table:nustardata} also includes  information about simultaneous observations performed with \xrt, if available.

\begin{table*} 
\caption[]{\textit{NuSTAR} observations of \obj.}
\centering 
\begin{tabular}{c|c|c|c |c|c|c}
\hline
\hline
Int &  \textit{NuSTAR} obsIDs  & duration (s) & date   &\textit{Swift}-XRT  obsIDs &  duration (s)  & date \\
(1) &  (2)& (3) & (4) & (5) & (6) & (7)   \\
\hline
N1  & 60001001001  &  385    & 2012-12-11 13:54:07  & 00030720050  & 7935  & 2012-12-11 05:48:59 \\
N2  & 60001001002  &  21891 & 2012-12-11 14:36:07  & 00030720050  & 7935  & 2012-12-11 05:48:59    \\
N3  & 60501024002  &  212202  & 2019-09-14 05:36:09  & 00095336001-00095336006   & 6482  &  2019-09-14 (6 obs)   \\
N4  & 90601630002  &  30725 & 2020-10-11 14:06:09 &  00034748029  & 2987 & 2020-10-11 19:23:51\\
N5  & 90701628002  & 20189  & 2021-08-16 02:06:09 &  00034748088  & 1531 & 2021-08-16 17:41:36 \\
N6  & 60701036002  & 22641   & 2022-05-06 08:51:09  &  $-$  & $-$  & $-$  \\
N7  & 90801633002  & 20842   & 2022-11-28 20:51:09  &  00096990017   &  826 &       2022-11-29 05:13:35  \\
\hline
\end{tabular}
\tablefoot{The columns are as follows: (1-2) Short name of the interval and the \textit{\nus}\ ID of observation.   (3) Duration of \textit{\nus}\ observation (in seconds).   (4) Date of \textit{\nus}\ observation.    (5) ID of  \xrt\ observations simultaneous to the \textit{\nus}\ one.    (6) Duration of \xrt\ observation (in seconds).  (7)  Date of \xrt\ observation.  }

 \label{table:nustardata}
\end{table*}

\section{Long-term X-ray light curve of BL\,Lacertae}\label{longlc}
\begin{figure*}
\centering{\includegraphics[width=0.99\textwidth]{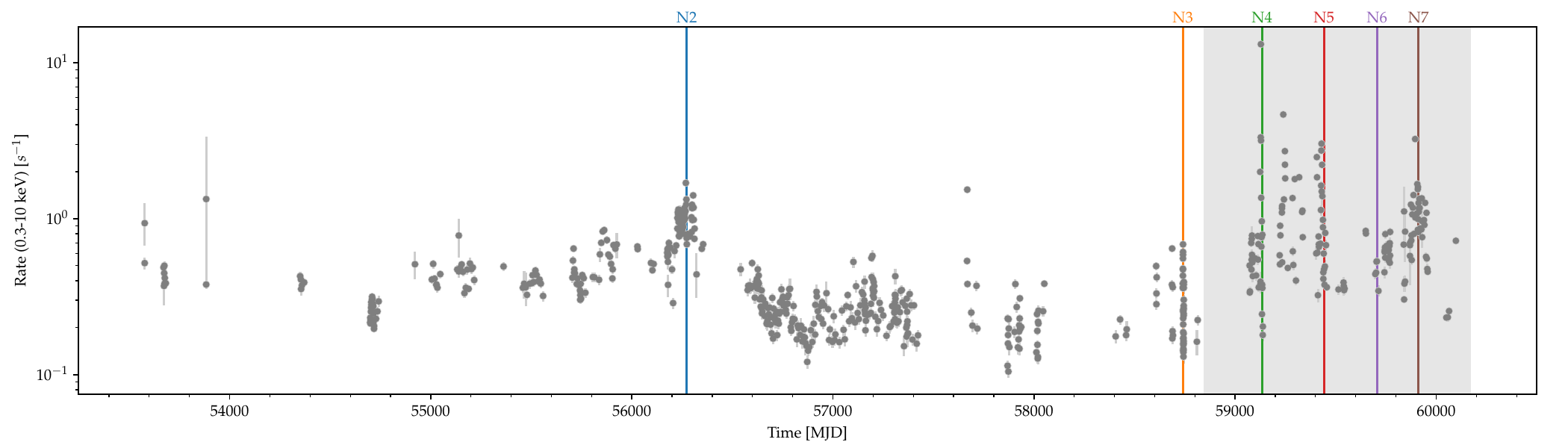}}
\caption{Long-term (2005-2023) X-ray light curve of \obj\ in the energy range of 0.3-10\,keV, including all \xrt\ observations of \obj. The shaded area indicates 2020-2023 observations. The coloured vertical lines indicate \textit{\nus}\ observations of \obj\ as discussed in Sect.\ref{nus}.}
\label{fig:lc_counts}
\end{figure*}

Figure~\ref{fig:lc_counts} presents the long-term light curve of \obj\ in the energy range of 0.3-10\,keV based on all \xrt\ observations performed between 2005 and 2023.  
The light curve presented in Fig.\,\ref{fig:lc_counts} demonstrates  significant temporal variability of the source. 
In particular, the average count rate of the entire period of observations is 0.5$\pm$0.1\,counts\,s$^{-1}$, while between 2020 and 2023 (the shaded area in Fig.\ref{fig:lc_counts}) this value is 1.0$\pm$0.1\,counts\,s$^{-1}$. 
The observations performed before 2020 are characterised by a count rate of 0.4$\pm$0.1\,counts\,s$^{-1}$. 
This indicates significantly higher fluxes in the period of 2020-2023 compared to previous years, that is, higher by a factor of 2. 
The highest point in the X-ray light curve is significantly above the others with the count rate being about 12\,counts\,s$^{-1}$. 
That corresponds to the \xrt\ observation performed on October 6, 2020. 
We note here that no intra-observation variability was detected  in this extraordinary observation of \obj.

\section{Spectral and temporal variability}\label{variability}
\subsection*{X-ray observations}
Between 2020 and 2023, \obj\ was observed more than 160 times with \xrt\ in the energy range of 0.3-10\,keV   (ObsIDs of 00030720231-00096565004).
In order to characterise all 2020-2023 observations in terms of spectral and temporal properties, we analysed each of these observations as described above, and fitted them using two spectral models: a single \po\ model and a log-parabola, both with the Galactic column density value of \nh=3.03 $\cdot$ 10$^{20}$\, cm$^{-2}$ as provided by \cite{Willingale13}.

The models mentioned above are as follows:  

\begin{itemize}
 \item a single power-law model, defined as
 \begin{equation}
\frac{dN}{dE}=N_p  \left( \frac{E}{E_0}\right)^{-{\gamma}},
\end{equation}

\item a logarithmic parabola (or curved power law), defined as
 \begin{equation}
\frac{dN}{dE}=N_l  \left( \frac{E}{E_0}\right)^{-({\alpha+\beta \log (E/E_0)})}.
\end{equation}
 \end{itemize}
The \po\ model is characterised by the normalisation $N_p$ and the spectral index $\Gamma$ (at energy  $E_0$), while the \lp\ model is described with the normalisation $N_l$ , the curvature parameter $\beta,$ and the spectral index $\alpha$ (at energy  $E_0$). 
In both cases, the scale energy $E_0$ is fixed at 1\,keV. 
The goodness of both fitted models is compared using the F-test \citep[e.g.][]{bevington2003data}.
We note here that  \lp\ models are only preferred when $\beta$ is negative, that is, a spectral upturn is observed in the range of the 0.3-10\,keV band.

\begin{figure*}
\centering{\includegraphics[width=0.99\textwidth]{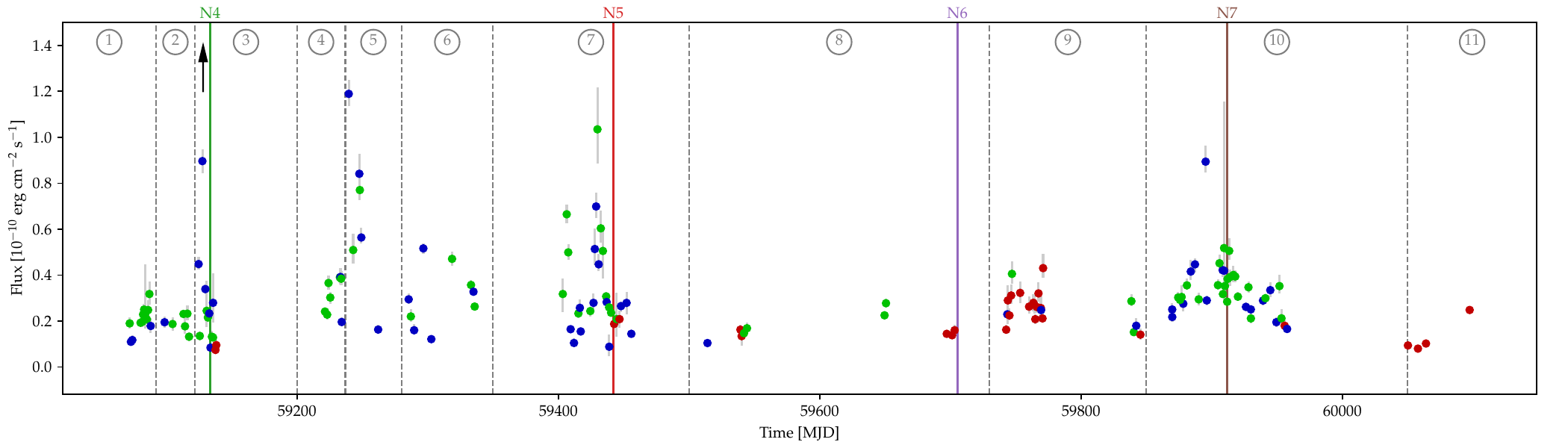}}
\caption{Long-term light curve of \obj\ presenting the 2020-2023 monitoring of the source in the energy range of 0.3-10\,keV.  Green, red, and blue points indicate different spectral shapes as defined in Sect.\,\ref{variability}. The vertical lines and numbers indicate the intervals studied in Sect.\ref{specvar}. Intervals N4-7 marked with coloured vertical lines correspond to the same intervals as those marked in Fig.\ref{fig:lc_counts}.}
\label{fig:lc}
\end{figure*}

Figure \ref{fig:lc} presents a  light curve of \obj\ in the energy range of 0.3-10\,keV for the period of 2020-2023. 
Different coloured points are used to denote the preferred spectral model used to describe the spectrum of a single observation, as follows:
\begin{enumerate}
 \item green points show spectra for which the \lp\ model with $\beta<$0 is preferred;
 \item red points show spectra for which the \po\ model is preferred and $\gamma<$2;
 \item blue points show spectra for which the \po\ model is preferred and $\gamma>$2.
\end{enumerate}

Based on the coloured SED, (1) refers to  the X-ray spectrum covering both the synchrotron and inverse Compton components. 
While cases (2.) and (3.) correspond to the situation when the X-ray spectrum dominates either the inverse Compton or synchrotron component only. 
As shown in Fig.\,\ref{fig:lc},  the curved power-law model with a negative curvature is the best description of more than half of the X-ray spectra. 
This spectral shape confirms the presence of both synchrotron and inverse Compton components in the energy range studied. 
The red coloured spectra, which describe the inverse Compton spectrum only, are in the minority and are only preferred for the low X-ray fluxes of \obj.

The highest point in the light curve corresponds to the flux of (3.32$\pm$0.04) $\cdot$ 10$^{-10}$ erg\,cm$^{-2}$s$^{-1}$ and this is the largest X-ray flux measure with XRT so far for \obj.

\begin{figure}
\centering{\includegraphics[width=0.49\textwidth]{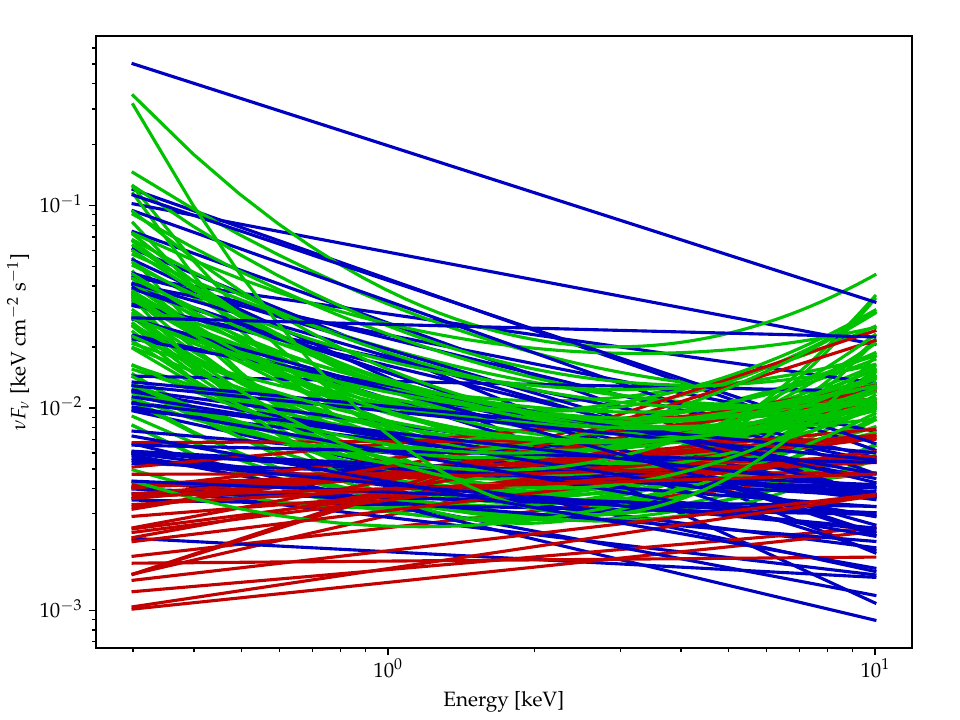}}
\caption{Spectral energy distributions for different activity states of \obj. Each spectrum represents one \xrt\ observation of the blazar. 
The same colour coding as in Fig.\ref{fig:lc} is used. }
\label{fig:seds}
\end{figure}

The models \po\ and \lp) used to describe the spectral shapes of \xrt\ observations are shown in Fig.\,\ref{fig:seds}.
The same colour coding is used as in the case of Fig. \ref{fig:lc}. 

\begin{figure}
\centering{\includegraphics[width=0.48\textwidth]{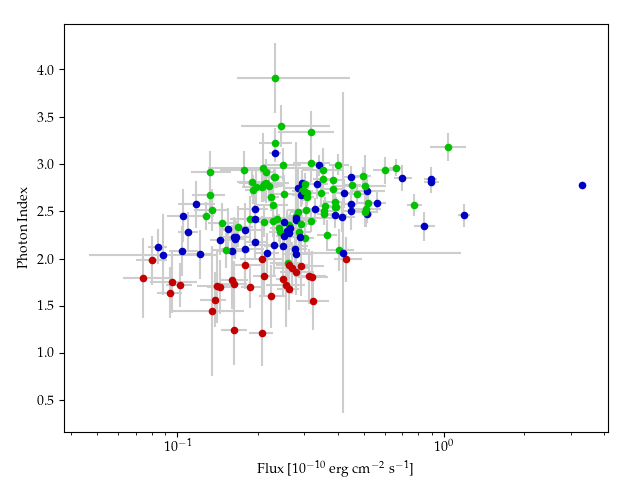}}
\caption{Comparison of the X-ray flux in the energy range of 0.3-10\,keV and the corresponding photon index for all 2020-2023 observations of \obj.
}
\label{fig:index_flux}
\end{figure}

Figure\,\ref{fig:index_flux} shows the relation between the  0.3-10\,keV flux measured for each XRT observation of \obj\ and the corresponding photon index: $\gamma$ or $\alpha$ for a \po\ or \lp\ fit, respectively. 
In order to check the relation between these two quantities, the Pearson correlation coefficient is calculated for all data points and for a subset of blue points only. 
The results are  0.40$\pm$0.04 and 0.54$\pm$0.10, respectively, indicating a lack of
correlation for the entire dataset, but a weak correlation when only a subset is taken into account.

\subsection*{Ultraviolet and optical observations}

\begin{figure}
\centering{\includegraphics[width=0.49\textwidth]{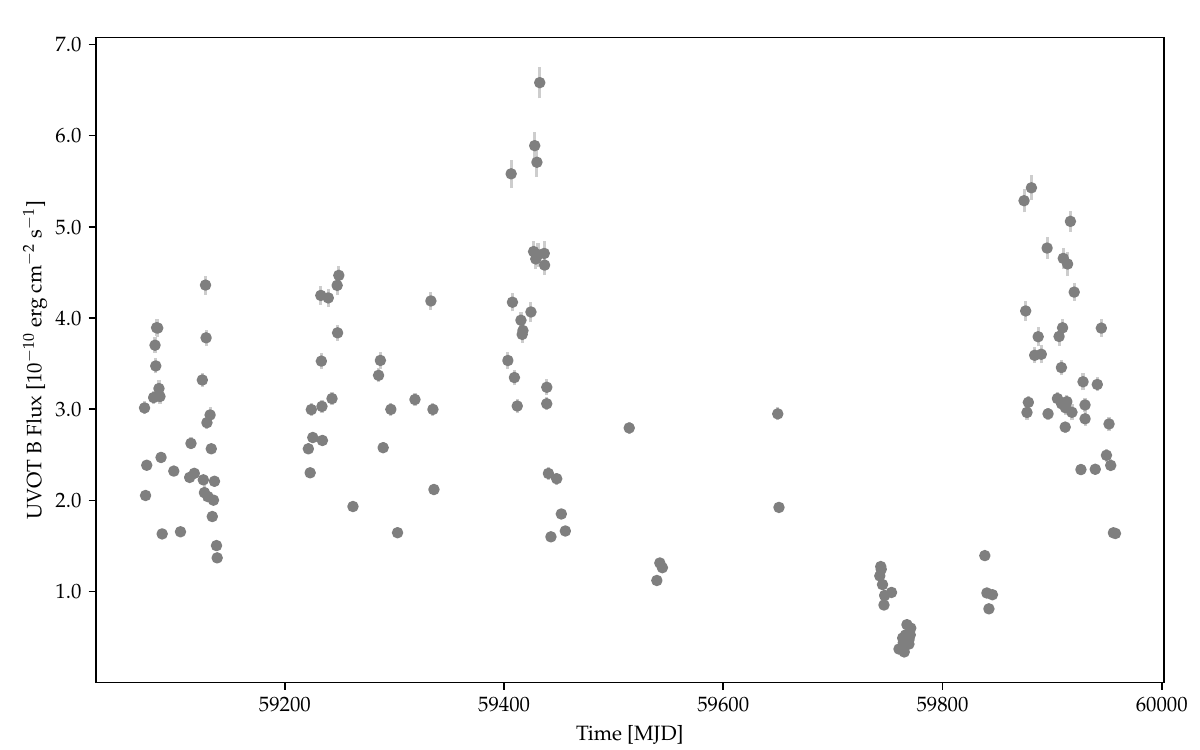}}
\caption{Long-term light curve of \obj\ presenting the 2020-2023 monitoring of the source in the optical B band.}
\label{fig:uvot_xrt_lc}
\end{figure}

\begin{figure}
\centering{\includegraphics[width=0.49\textwidth]{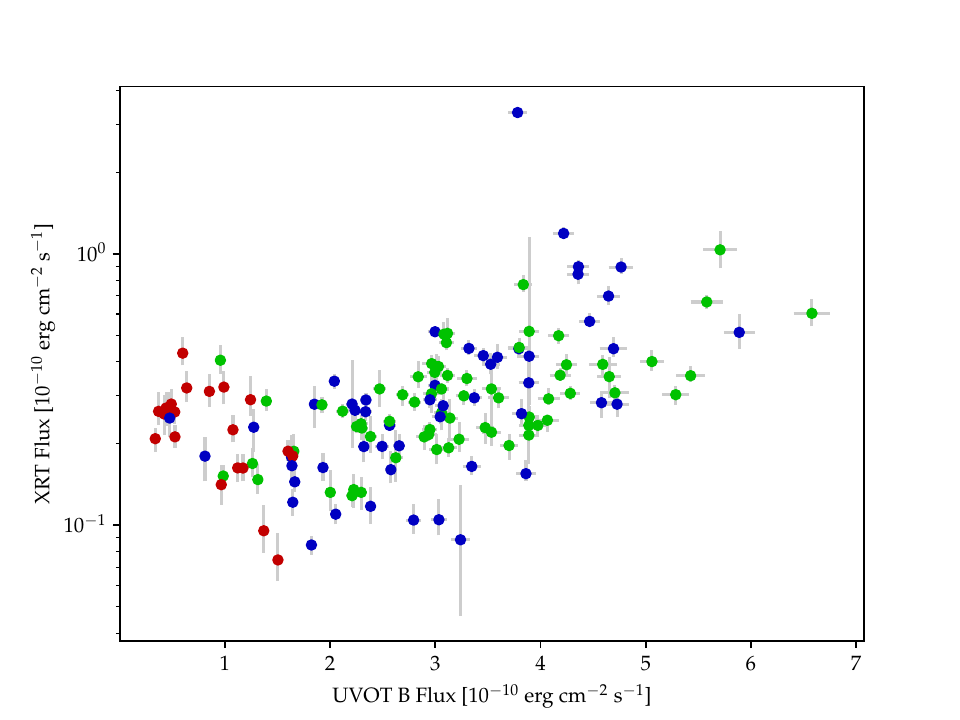}}
\caption{Comparison of the optical and corresponding X-ray fluxes. Plot based on the data presented in Fig.\ref{fig:lc} and Fig.\,\ref{fig:uvot_xrt_lc}. }
\label{fig:uvot_xrt}
\end{figure}

Together with X-ray observations, \textit{Swift} observed \obj\ simultaneously in the UV and optical wavebands. 
Figure\,\ref{fig:uvot_xrt} presents a comparison of X-ray and optical fluxes in the energy range of 0.3-10\,keV and the B band. 
Each data point corresponds to one observation.
Significant variability  is seen in both energy regimes, as presented in the light-curve plots (see Figs.\ref{fig:lc} and \ref{fig:uvot_xrt_lc} for  reference).
The flux changes seen in the X-ray and optical band, however, are not identical.
Figure\,\ref{fig:uvot_xrt}  also shows that two distinct trends are present. 
The X-ray and optical fluxes are correlated for high optical fluxes above 2 $\cdot$ 10$^{-10}$ erg\,cm$^{-2}$s$^{-1}$, and the opposite trend is present below this value.
The exceptional X-ray activity observed on  October 6, 2020, has no counterpart in the optical range.

%----------------------------
%----------------------------

\subsection*{Spectral variability} 
A spectral power-law index ($\Gamma_{syn}$) describing a low-energy bump of the broadband SED  is  calculated between an optical B point and the low-energy end of the X-ray spectrum; that is, the X-ray model flux at 0.3 keV. Here,  $\Gamma_{syn}$ is defined as follows: 

\begin{equation}
 \Gamma_{syn} = \frac{\log F{_X} - \log F{_O}}{\log E{_X} - \log E_O},
\end{equation}
where  E${_X}$ is the energy of the X-ray point, E${_O}$ is the energy of an optical point, and F$_X$ and F$_O$ are X-ray and optical fluxes, respectively.
We note here that E${_X}$ = 0.3\,keV  , while E${_O}$ = 2.85\,eV.
Also, $F{_X}$ and $F{_O}$ are normalised by keV\,$cm^{-2}$ s$^{-1}$, while   $E{_X}$ and  $E{_O}$ are normalised by keV.

\begin{figure}
\centering{\includegraphics[width=0.49\textwidth]{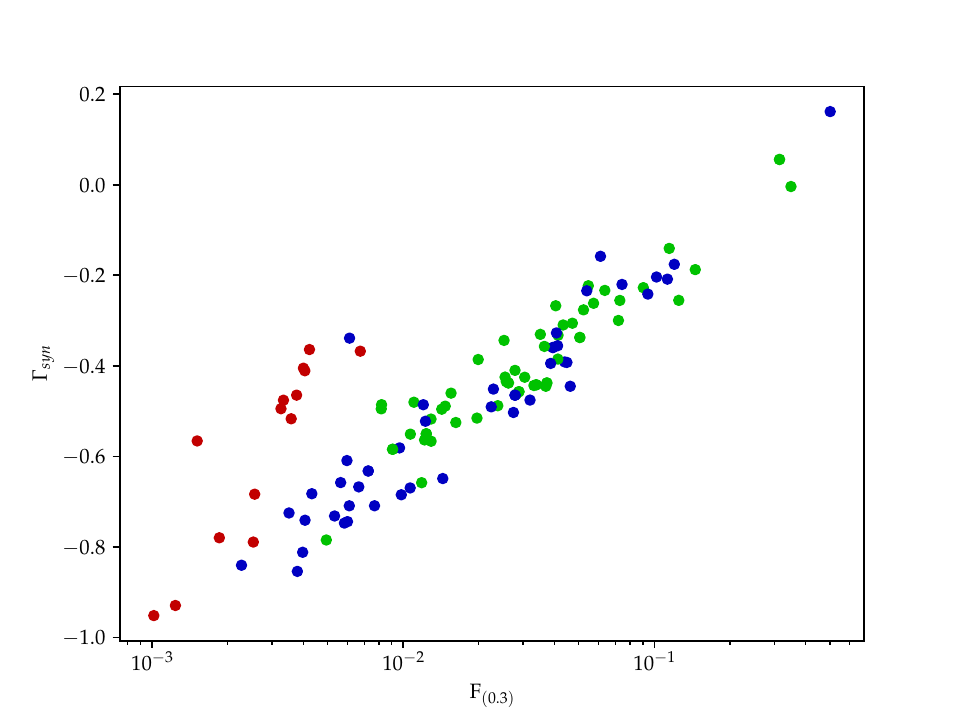}}
\caption{Comparison of $\Gamma_{syn}$ as a function of the X-ray model flux at 0.3\,keV. The colour-coding (same  as in Fig. \ref{fig:lc}) denotes the different models describing the spectral shapes.}
\label{fig:gamma_f03}
\end{figure}

Figure\,\ref{fig:gamma_f03} shows a comparison between  the $\Gamma_{syn}$ index and the X-ray flux at 0.3\,keV.
 The same colour coding as in Fig.\ref{fig:lc} is used to indicate the preferred spectral models of the single X-ray spectra, and the green colour corresponds to the curved spectra with a negative curvature parameter, while the power-law spectra are denoted with red and blue points. 
A clear linear trend is present for blue and green points, while the outliers in the plot are all denoted with red points. 
This indicates that $\Gamma_{syn}$ is correlated with the X-ray flux in the case where the X-ray spectrum covers mainly or only the synchrotron part of the SED. 
Red points ---which do not follow the trend--- fully describe the high-energy part of the SED. 
Interestingly, $\Gamma_{syn}$ is not correlated with the X-ray model flux at 10\,keV.

\begin{figure}
\centering{\includegraphics[width=0.49\textwidth]{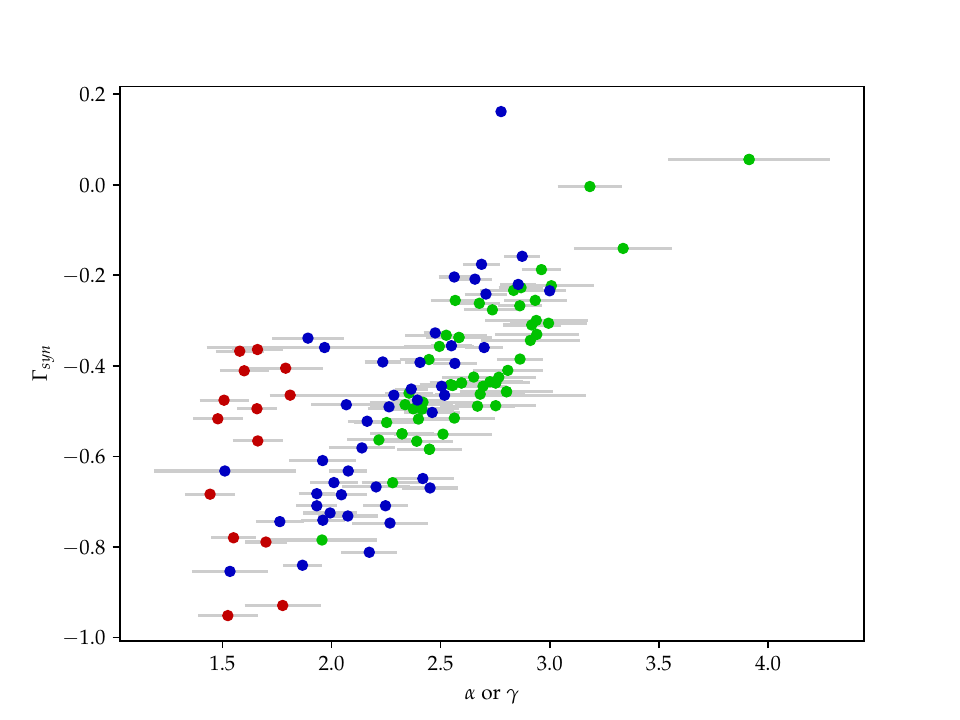}}
\caption{Comparison of $\Gamma_{syn}$ as a function of X-ray spectral index  ($\gamma$ or $\alpha$ depending on the spectral model used). The colour coding (same  as in Fig.\ref{fig:lc}) denotes the different models describing the spectral shapes.}
\label{fig:gamma_vsalpha}
\end{figure}

A comparison of $\Gamma_{syn}$ as a function of the X-ray photon index describing the \xrt\ spectrum is shown in Fig.\,\ref{fig:gamma_vsalpha}.
The spectral index used here is either $\gamma$ or $\alpha$  depending on the spectral model preferred to describe the X-ray spectrum of \obj. 
A linear trend is present for green and blue points, while the red points do not follow this relation, which is similarly to in the case of Fig.\,\ref{fig:gamma_f03}.
Again, this relation linking $\Gamma_{syn}$ and the X-ray photon index is not followed by the red points but is followed by a few of the blue points.

\section{Time-resolved spectral variability}\label{specvar}

To investigate the spectral variability in greater detail, the long-term X-ray light curve of \obj\ was divided into shorter intervals characterised by similar flux levels and consisting of a sufficient number of observations to constrain spectral parameters with accuracy and allow investigation of their evolution. 
The division into ten intervals is presented in Fig.\,\ref{fig:lc} and the intervals are numbered for clarity. 
In the figure, interval 11 is also marked. 
For this subset of observations, the X-ray spectrum is well described with a \po\ model with a photon index of 1.82$\pm$0.04.

\begin{figure}
\centering{\includegraphics[width=0.49\textwidth]{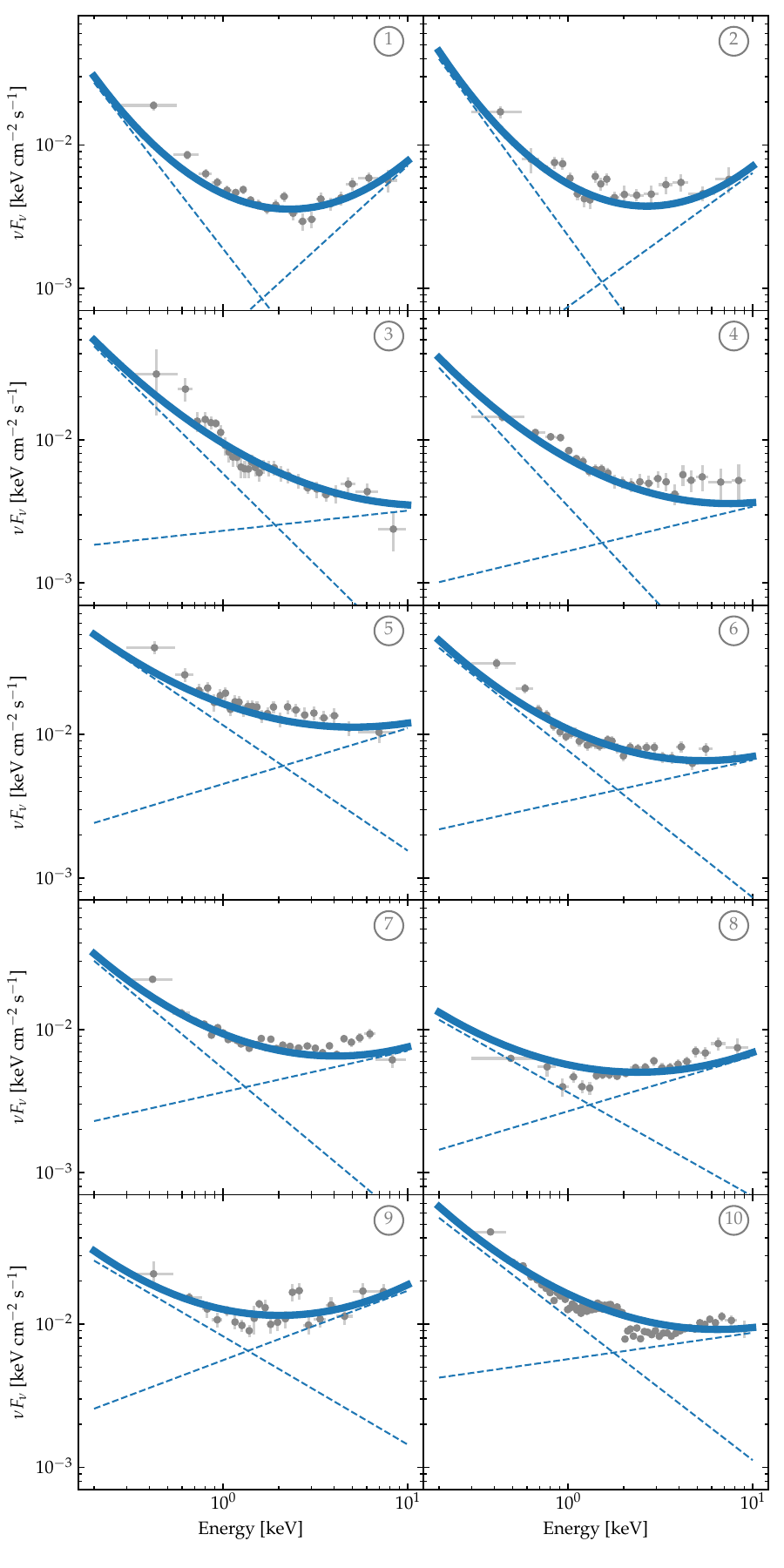}}
\caption{Spectral variations during the observation period. Each spectrum is derived for the interval defined in Fig.\,\ref{fig:lc}.
Two spectral models, namely a log-parabola and double power law, as well as data points are shown in each plot. 
The spectral parameters of each plot are given in Table\,\ref{table_2comp}.}
\label{fig:components}
\end{figure}

The spectral analysis of each of the intervals selected includes spectral fitting with a  \lp\ model and a double power-law model. 
In both cases, the Galactic absorption was also included in the spectral model.
Table\,\ref{table_2comp} and Fig.\,\ref{fig:components} present the variations of spectral parameters derived for the spectral fits. 

The photon index of the low-energy component changes between 2.7 and 3.6, while that of the high-energy component varies between 0.8 and 1.8. 
Also, the crossing point of two spectral components moves to higher energies with increasing total flux.
However, the range of detected values of the crossing point of 1.3-2.1\,keV is small. 

The relative contribution of the synchrotron component, which is 12$-$45$\%$ depending on the activity state of \obj,\ shows that the X-ray spectrum of \obj\ is dominated by the high-energy (inverse Compton) component. 
The contribution of the low-energy component is the largest for interval 3, which is the epoch where the source is in its highest state.

The separation of synchrotron and inverse Compton components allows a comparison of the variability in the two components.

A first test investigating whether the low-energy end of the inverse Compton component reveals any significant variability within the data set. 
For this purpose, the low-energy component  a fixed value of a photon index and normalisation is used and the parameters of the high-energy component are left free.
Different values of the photon index and normalisation were tested, including all the possibilities listed in Table\,\ref{table_2comp}.
Such a spectral fitting in all cases, however, resulted in reduced $\chi^2$ values of significantly higher than 1, and large residuals at the high-energy end of the spectrum, discrediting such a scenario.

\begin{table*}  
\caption[]{Results of the time-resolved spectral analysis of the \textit{Swift} data.}
\centering
\begin{tabular}{c|c|c|c|c|c |c|c |c}
\hline
\hline
 Interval & $\Gamma{_1}$ & $\Gamma{_2}$ & E$_{cross}$ &  F$_{low}$ &  F$_{high}$ & Synchrotron contribution & $\alpha$ & $\beta$    \\ 
           (1) &  (2) & (3) & (4) &  (5) &  (6) & (7)  & (8)  &  (9)  \\
\hline 
1 &  3.65$\pm$0.05      & 0.82$\pm$0.03 &       1.62    &       0.37$\pm$0.01   &                 1.26$\pm$0.03   &   23$\%$  &  2.59$\pm$0.06  &  -0.82$\pm$0.01   \\
2 &  3.76$\pm$0.06      & 1.06$\pm$0.03 &       1.52    &       0.39$\pm$0.01   &                 1.28$\pm$0.03   &   23$\%$  &  2.73$\pm$0.05  &  -0.85$\pm$0.01    \\
3 &  3.27$\pm$0.04      & 1.85$\pm$0.04 &       1.91    &       0.80$\pm$0.02   &                 1.10$\pm$0.02   &   45$\%$  &  2.78$\pm$0.03  &  -0.35$\pm$0.01     \\
4 &  3.38$\pm$0.05      & 1.68$\pm$0.03 &       1.52    &       0.52$\pm$0.01   &                 1.97$\pm$0.01   &   21$\%$  &  2.72$\pm$0.06  &  -0.42$\pm$0.01     \\
5 &  2.87$\pm$0.04      & 1.61$\pm$0.04 &       2.10    &       1.77$\pm$0.03   &                 2.75$\pm$0.03   &   39$\%$  &  2.46$\pm$0.05  &  -0.33$\pm$0.01     \\
6 &  3.02$\pm$0.05      & 1.71$\pm$0.04 &       1.85    &       1.01$\pm$0.02   &                 1.79$\pm$0.02   &   36$\%$  &  2.60$\pm$0.04  &  -0.41$\pm$0.01     \\
7 &  3.08$\pm$0.05      & 1.71$\pm$0.05 &       1.32    &       0.52$\pm$0.01   &                 2.11$\pm$0.02   &   12$\%$  &  2.51$\pm$0.04  &  -0.42$\pm$0.01     \\
8 &  2.73$\pm$0.05      & 1.62$\pm$0.03 &       1.31    &       0.31$\pm$0.01   &                 1.67$\pm$0.01   &   16$\%$  &  2.27$\pm$0.04  &  -0.36$\pm$0.01     \\
9 &  2.76$\pm$0.08      & 1.51$\pm$0.07 &       1.36    &       0.30$\pm$0.01   &                 2.45$\pm$0.02   &   11$\%$  &  2.27$\pm$0.04  &  -0.45$\pm$0.01     \\
10&  3.00$\pm$0.08      & 1.81$\pm$0.06 &       1.76    &       1.11$\pm$0.03   &                 2.44$\pm$0.01   &   31$\%$  &   2.61$\pm$0.03  &  -0.38$\pm$0.01     \\
\hline
\hline
\end{tabular}
\tablefoot{The columns are as follows: (2)-(3) photon indices for a double-power-law fit with Galactic absorption; (4) energy value for the crossing point in keV of two power-law lines; (5)-(6) flux of the low- and high-energy components in $10^{-11}\,erg\,cm^{-2}\,s^{-1}$; (7) relative contribution of the synchrotron component; (8) photon index of a \lp\ fit; (9) curvature parameter of a \lp\ fit.}
\label{table_2comp}
\end{table*}

\section{Hard X-ray observations of \obj\ with \textit{\nus}}\label{nus}
Additional information about X-ray spectra of \obj\ can be obtained thanks to hard X-ray observations performed with \textit{\nus}. 
As listed in Table \ref{table:nustardata}, there are six observations of \textit{\nus}\ with exposure longer than 1ks. 
For five of them, there are also simultaneous XRT observations, allowing us to constrain the X-ray emission of the blazar in the energy band starting from 0.3 up to ~70keV. 

In the case of all six \textit{\nus}\ and simultaneous \xrt\ observations, both models: a single  \po\ and  \lp\ were used. The preferred model was selected using the  F-test. 
Based on the test for all cases, the X-ray emission of \obj\  is best represented with a curved \po\ model in four cases (N2, N3, N4, and N7) and a single  \po\ model in the others.
Where a \po\ model is preferred, the hard X-ray spectrum is only able to clearly describe the inverse Compton part of the spectrum, and there is no indication of the presence of a spectral upturn in this energy regime. 
We also note here that, for the N4 epoch, the curvature seen in the spectrum is marginal and a point of spectral upturn is also not clearly indicated. 
In the case of epochs N2, N3,  and N7 \xrt-\textit{\nus}\ joint spectra describe both spectral components of \obj. 
However, all three epochs differ in terms of spectral curvature and the position of the crossing point. 
For N2 and N7, a crossing point is located above 2\,keV and a curvature parameter is between (-$0.2$) and (-$0.1$), while in the case of N3, the spectral curvature is more significant, with $\beta$=$-$0.5, and the crossing point is at E=8\,keV. 
The N3 epoch corresponds to rather low activity of the blazar (see Fig.\ref{fig:lc_counts}). Thus, the location of the crossing point for this epoch is significatly higher than  the crossing points observed for the \xrt\ observations.

\begin{figure}
\centering{\includegraphics[width=0.49\textwidth]{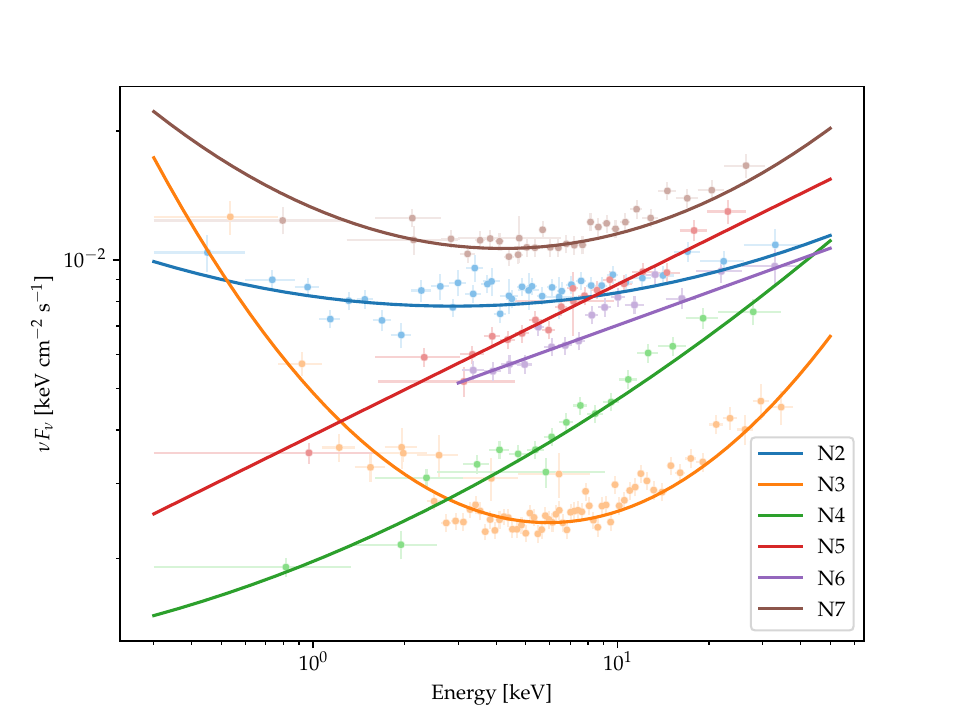}}
\caption{The models of  spectral energy distributions of \obj\ obtained based on \textit{\nus}\ and \xrt\ simultaneous observations. Different colours denote separate epochs of observations as defined in Table\ref{table:nustardata}. Data points for each set of observations are denoted with the same colours.} 
\label{fig:nustar_sed}
\end{figure}

\begin{table}  
\caption[]{Spectral fitting of the \textit{\nus}\ observations of \obj.
}
\centering
\begin{tabular}{c|c|c|c}
\hline
\hline
 Interval & Normalisation  & $\alpha$ or $\gamma$ & $\beta$    \\
           (1) &  (2) & (3) & (4)  \\
\hline 
N2  &   8.21$\pm$0.17   &       2.10$\pm$0.03   &       $-$0.11$\pm$0.03  \\
N3  &   4.88$\pm$0.13   &       2.79$\pm$0.03   &       $-$0.51$\pm$0.02  \\
N4  &   1.97$\pm$0.12   &       1.71$\pm$0.06   &       $-$0.1$\pm$0.03  \\
N5  &   3.89$\pm$0.26   &       1.65$\pm$0.07   &       $-$     \\      
N6  &   3.88$\pm$0.06   &       1.74$\pm$0.12   &       $-$     \\      
N7  &   13.22$\pm$0.72  &       2.30$\pm$0.05   &       $-$0.24$\pm$0.03  \\
\hline
\hline
\end{tabular}
\tablefoot{The columns are as follows: (1) Name of interval. (2) Normalisation given in $10^{-3}\,$keV$\,$cm$^{-2}\,$s$^{-1}$. (3) Photon index $\alpha$ or $\gamma$. (4) Curvature parameter $\beta$.}
\label{table_nustarfits}
\end{table}

\section{Summary}\label{summary}
In this work, we present an analysis of long-term X-ray and optical observations of the blazar \obj. 
Numerous observations of the source performed with \xrt, \uvot, and \textit{\nus}\ revealed different activity states of this AGN.

Studies of X-ray, optical, and ultraviolet observations of \obj\ have confirmed the strong temporal variability of the source, which is associated with spectral changes. 
Based on the long-term X-ray observations collected with \xrt, the difference between the lowest and the highest count rates observed is 12\,cts\,s$^{-1}$, which is 24 times more than the average count rate. 
Large changes in the flux are also present in the optical and ultraviolet domains. 
The highest X-ray point reported, which corresponds to the observations performed on October 6, 2020, represents the historical maximum of the \xrt\ monitoring of \obj.
However, we note that the optical flux measured simultaneously in X-rays corresponds to a high state of the source, but does not represent the historical optical maximum.

Our analysis of the long-term behaviour of \obj\ based on \xrt\ observations collected in 2006-2023 shows that the largest activity of the blazar started in 2020 and continued in the following years. 
Our detailed analysis of the X-ray observations collected from 2020 to 2023 with \xrt \ reveals significant spectral and temporal variability of the source, and a complex behaviour during this period of intensive observations.
Similarly, in the optical frequencies, the 2020-2023 fluxes are significantly higher than in the previous years \citep[see e.g. the optical monitoring of BL\,Lacertae with the ATOM telescope][]{Wierzcholska_atom}.

The large set of X-ray observations performed with the two X-ray instruments \xrt\ and \textit{\nus}\ allows simultaneous measurements of the high- and low-energy ends of the particle distribution function for this source. 
In the X-ray energy range of 0.3-10\,keV, significant variability can be seen both in the synchrotron and inverse Compton branches in terms of amplitude and variability scales.
The amplitude of this variability is similar for synchrotron and inverse Compton ranges.
Given the similar variability scales observed in both the synchrotron and inverse Compton branches, 
radiative cooling does not dominate the variability characteristics of \obj.

The spectral model describing single X-ray observations is either a \po\ or \lp\ model, depending on the epoch of the observation. 
The case of the curved model corresponds to the situation where the transition between the synchrotron and inverse Compton component is visible in the energy range of 0.3-10\,keV. 
The spectral curvatures described by the $\beta$ parameter in the model range between $-$0.1 and $-$0.8 indicate changes in the slope of the spectral components. 
For the curved \po\ model, for all cases, the spectral upturn is located below 2\,keV. 
However, given the variety of the X-ray spectra, which exhibit either synchrotron or inverse Compton or both spectral components with the spectral upturn, we note that the crossing energy for \obj\ can be outside the energy band of 0.3-10\,keV. 
In particular, for the \po\ spectra with a photon index of smaller than 2, the spectral upturn is located below 0.3\,keV, while for the \po\ spectra with a photon index larger than 2, the spectral upturn is located above 10\,keV. 

Given this characteristic, \obj\ is a unique example of a source exhibiting features of HBL-, IBL-, and LBL-type blazars, depending on the epoch of the observation. 
This occurs because the photon index and curvature of the spectrum change between epochs.
The X-ray spectrum of \obj\ can either cover the synchrotron branch of the broadband SED, the inverse Compton one, or both, which correspond to the HBL-, LBL-, or IBL-type blazar characteristics, respectively. 
Interestingly, in the low state of the source, the blazar exhibits an X-ray spectrum with an LBL-like character, while for higher fluxes the source shows either HBL-like or IBL-like features  in its  X-ray spectrum. 

The IBL-like characteristic of \obj\ confirms that the energy range of 0.3-10\,keV is a region where two spectral components, synchrotron and inverse Compton, meet each other, which was already reported in previous studies focusing on \obj.
These studies reported the spectral transition of the synchrotron and inverse Compton component at energies below 2\,keV.
In particular, \cite{Tanihata00} reported the spectral break at about 1\,keV.
\cite{Donato05} found the transition point of the spectral components at 0.5\,keV.
The recent studies by \cite{Wierzcholska_swift} showed that regardless of the epoch of observation, the spectral break is located at 1\,keV. At the same time, both spectral components change in slope and flux normalisation. 
In the present work, we report similar values of the crossing point based on \xrt\ observations of synchrotron and inverse Compton components ranging between 1.3 and 2.0\,keV and being almost constant within the uncertainties, regardless of significant changes in flux and photon index.

A two-component model was also preferred for joint XMM-Newton-\textit{\nus}\ observations of the source, with a crossing point located at approximately 2\,keV \citep{Peirsonixpe}.
The simultaneous IXPE observations during this campaign resulted in the first  detection of X-ray polarisation in \obj, which the authors found to be possibly related to the larger contribution of the
synchrotron component  to the X-ray spectrum of the blazar. 
The X-ray polarisation was not detected in the previous, low-state observations of \obj\ with IXPE \citep{Middei}.

To characterise spectral changes in the low-energy end of the electron distribution function, we used the $\Gamma_{syn}$ parameter, which we calculated for each individual X-ray point during the 2020-2023 epoch.
A comparison of $\Gamma_{syn}$ and X-ray flux at 0.3\,keV revealed a linear relation for those spectra, which correspond to mainly the synchrotron component, except in the case of the X-ray slopes, which characterise the inverse Compton part of the SED.
Also, a comparison of $\Gamma_{syn}$ as a function of the X-ray photon index revealed a similar feature. A linear trend is present for all points that correspond to spectra representing synchrotron components.
These comparisons confirm that \obj\  is a blazar for which the X-ray band of 0.3-10 keV can correspond to different bumps on the SED.

The comparison of the X-ray flux and corresponding photon index revealed that the harder-when-brighter relation is not present at any point during the entire set of 2020-2023 observations. 
The colour coding used revealed that spectra with different characteristics (different colours) are clustered in the figure.

The dense monitoring of \obj\ performed with \textit{Swift} allowed us to perform time-resolved spectral analysis, focusing on intervals of the blazar's activity.
These studies confirm the location of the spectral upturn in the \xrt\ band. 
The analysis of time-resolved spectra also revealed strong temporal variability associated with spectral changes.
The position of the crossing point varies with temporal changes but remains confined to a small range of energies.
We reject the hypothesis that only one spectral component varies with time while the other remains constant.
Furthermore, in the case of the highest X-ray fluxes observed, the largest contribution of the low-energy component during these observations supports the scenario according to which the flaring activity of \obj\ is mainly caused by the high-energy tail of the synchrotron component. 
A similar trend was reported in the case of another IBL-type blazar, S5\,0716+714 \citep{Ferrero06}.

Different activity states of \obj\ were also observed with \textit{\nus}. 
These hard X-ray observations of \obj\ confirm that its X-ray spectrum exhibits different  properties, 
including the situation where it covers only the inverse Compton bump, and the presence of a crossing point in the hard X-ray regime.  
The concave curvature of the \textit{\nus}-\xrt\ spectrum of the blazar was well constrained and there was no need for an additional absorption component below 2\,keV. 
We also note here that in the case of two observations performed with \textit{\nus}, a crossing point is located at the energy of about 2\,keV, while in one case the upturn is at about 8\,keV, which is the highest upturn energy reported for this blazar. 
\textit{\nus}\ observations of \obj\ do not support the hypothesis that a spectral crossing point is moving to higher energies with increasing flux. 

We note here, however, that for the X-ray spectra provided by the integrated \xrt\ observations, the upturn point changes its position only in a limited range of energies and does not reach energies as high as in the case of \textit{\nus}\ spectra. 
This may be the result of the averaging of the long-term observations  effect.
We conclude that the energy crossing point moves to higher energies only for short periods of the observations. 
These higher energies of the spectral upturn are also not reproduced in the single-observation spectra of \xrt\ data because of insufficient temporal resolution.

\begin{acknowledgements}
The project is co-financed by the Polish National Agency for Academic Exchange.
This research has made use of the    \textit{NuSTAR} Data Analysis Software (NuSTARDAS) jointly developed by the ASI Space Science Data Center (SSDC, Italy) and the California Institute of Technology (Caltech, USA). The authors acknowledge the use of public data from the \textit{Swift} data archive.
The authors  gratefully acknowledge Polish high-performance computing infrastructure PLGrid (HPC Center: ACK Cyfronet AGH) for providing computer facilities and support within computational grant no. PLG/2023/016899.
\end{acknowledgements}

\bibliographystyle{aa} 
\bibliography{main.bib}
\end{document}